\documentclass{amsart}
\usepackage{amsfonts}

\newtheorem{theorem}{Theorem}
\theoremstyle{plain}

\newtheorem{corollary}{Corollary}

\newtheorem{definition}{Definition}
\newtheorem{example}{Example}

\newtheorem{lemma}{Lemma}

\newtheorem{proposition}{Proposition}

\numberwithin{equation}{section}

\begin{document}
\title[]{Tensor operators and Wigner-Eckart theorem for the quantum
superalgebra $U_{q}[osp(1\mid 2)]$
}
\author{Marek Mozrzymas}
\address{Institute of Theoretical Physics, University of Wroclaw, \\
pl. Maxa Borna 9, 50-204 Wroclaw, Poland}
\email{marmoz@ift.uni.wroc.pl}
\thanks{}
\date{8 April 2004}
\thanks{}

\begin{abstract}
Tensor operators in graded representations of $\ Z_{2}-$graded Hopf algebras
are defined and their elementary properties are derived. Wigner-Eckart
theorem for irreducible tensor operators for $U_{q}[osp(1\mid 2)]$ is
proven. Examples of tensor operators in the irreducible representation space
of Hopf algebra $U_{q}[osp(1\mid 2)]$ are considered. The reduced matrix
elements for the irreducible tensor operators are calculated. A construction
of some elements of the center of $U_{q}[osp(1\mid 2)]$ is given.
\end{abstract}

\maketitle

\section{Introduction}

\bigskip This article is a continuation of the study of the properties of
irreducible representations (so called Racah-Wigner calculus) of the quantum
superalgebra $U_{q}[osp(1\mid 2)]$. In previous papers \cite{1, 2, 3} it was
shown that it is possible to construct Racah-Wigner calculus for this
quantum superalgebra in a completely similar way as in the classical Lie
algebra $su(2)$ \cite{4} and the quantum algebra $U_{q}(su(2))$ \cite{5, 6,
7}. It is quite remarkable that \ all topics that are relevant for the
Racah-Wigner calculus for $su(2)$ or $U_{q}(su(2))$ have their direct
super-analogue in the representation theory quantum superalgebra $%
U_{q}[osp(1\mid 2)]$.

An important part of the classical Racah-Wigner calculus are definition and
properties of tensor operators in the representation spaces.The concept of
tensor operators is very important in applications of symmetry techniques
(Lie groups and algebras) in theoretical physics. The irreducible tensor
operators for the Lie group of space rotations were first introduced by
Wigner \cite{8}. Equivalent definition of tensor operators for the
coresponding Lie algebra was given by Racah\cite{9}. These tensor operators
play very important role in the theory of angular momentum in quantum
physics.

The importance of tensor operators in the representation theory of the Lie
groups and algebras leads naturally to investgate the concept of tensor
operator for quantum groups and algebras as well as for the quantum
superlagebras. The classical Wigner-Racah definition of the irreducible
tensor operator has been extented to the quantum lie algebras in papers \cite%
{6, 7, 10} and Wigner-Eckart theorem has been proved in the similar way as
in classical undeformed symmetry structures. In papers \cite{11, 12} a new,
more general definitions of tensor operators for arbitrary Hopf algebra has
been proposed. According these definitions tensor operators are
homomorphisms \ of some Hopf algebra representations. The new general
definitions, on one hand are equivalent to the classical Wigner-Racah
definitions if the corresponding Hopf algebra is $su(2)$ or $U_{q}(su(2))$,
on the other hand they allow to deduce easier general properties of tensor
operators from basic properties of Hopf algebra representations.
Wigner-Eckart theorem for irreducible tensor operators for Hopf algebras has
been proved in paper \cite{13}, where a more general version of the
definition from paper \cite{12} has been used.

In this paper we define tensor operator for $\mathtt{Z}_{2}$-graded Hopf
algebras in a similar way as in papers \cite{12, 13}. We study the basic
properties of the linear operators acting in the graded irreducible
representation spaces of the quantum superalgebra $U_{q}[osp(1\mid 2)]$. In
particular we prove Schur lemma for $U_{q}[osp(1\mid 2)].$ Next we fomulate
and prove Wigner-Eckart theorem for irreducible tensor operators of the
quantum superalgebra $U_{q}[osp(1\mid 2)]$. The proof is based on the
properties of $U_{q}[osp(1\mid 2)]$ representations, in particular a
conclusions from Schur lemma play important role in it. It is remarkable
that Wigner-Eckart theorem for $U_{q}[osp(1\mid 2)]$ has exactly the same
form as in the classical case \ $su(2)$ or $U_{q}(su(2))$ i.e. the matrix
elements of components of irreducible tensor operator for $U_{q}[osp(1\mid
2)]$ are proportional to Clebsch-Gordan coefficients and the proprtionality
coefficient (reduced matrix element) has the same properties that in case of
$su(2)$ or $U_{q}(su(2)).$ Using properties of representatins of \ graded
Hopf algebras we construct two classes of tensor operators for $%
U_{q}[osp(1\mid 2)].$ In the first class tensor operators act in the adjoint
and regular representations of $U_{q}[osp(1\mid 2)].$ The second class of
tensor operators consists of the irreducible tensor operators acting in the
irreducible representation spaces of $U_{q}[osp(1\mid 2)].$ As an
application of Wigner-Eckart theorem we calculate the reduced matrix
elements for the irreducible tensor operators. Finally we give a method of
constructing of elements of the center of $U_{q}[osp(1\mid 2)],$ based on
the properties of tensor product of irreducible representations.

This paper has the following structure. In Section II we give a rewiew of
basic definitions and properties of graded representations, we define tensor
operators for $\mathtt{Z}_{2}$-graded Hopf algebra and we give some examles
of tensor operartors. In Section III we rewiew basic properties of grade
star representations of $U_{q}[osp(1\mid 2)].$ Using these properties we
prove Schur lemma next we formulate and prove Wigner-Eckart theorem for the
quantum superalgebra $U_{q}[osp(1\mid 2)]$. In section IV we consider
examples of tensor operators for $U_{q}[osp(1\mid 2)]$, we calculate the
reduced matrix element for the irreducible ones and we give a construction
of some elements of the center of $U_{q}[osp(1\mid 2)]$.

\section{Tensor operators for $\mathtt{Z}_{2}$-graded Hopf algebras.}

\bigskip We begin by recalling the definition of the $\mathtt{Z}_{2}$-graded
Hopf algebra.

\begin{definition}
A $\mathtt{Z}_{2}$-graded Hopf algebra\textrm{\ }is a vector space \textrm{A
}over complex field $\mathbf{C}$\textrm{\ }such that \textrm{A }=$\oplus
_{\alpha \in \mathtt{Z}_{2}}$\textrm{A}$_{\alpha }$\textrm{. }The elements $%
\mathrm{a}$ of \textrm{A}$_{\alpha }$ are said to be homogenous of degree $%
\alpha $ ($\alpha =0\leftrightarrow $even, $\alpha
=1\leftrightarrow $ odd) and their degree will be noted $\deg
(\mathrm{a})\equiv \mid \mathrm{a}\mid \in \mathtt{Z}_{2}.$ We
assume that the unit $\mathbf{1}$ of a graded algebra belongs to
\textrm{A}$_{0}$. In the following all Greek indices will belong
to $\mathtt{Z}_{2}$. Further we have in \textrm{A}\\
%\begin{definition}
1) an associative multiplication, $m:\mathrm{A}\otimes \mathrm{A}\rightarrow
\mathrm{A}$, $m(\mathrm{A}_{\alpha }\otimes \mathrm{A}_{\beta })\subset
\mathrm{A}_{\alpha +\beta }$,

$m(\mathrm{a}\otimes \mathrm{b})=\mathrm{ab}$, \textrm{a,b}$\in $\textrm{A},%
\begin{equation*}
m\circ (id_{\mathrm{A}}\otimes m)=m\circ (m\otimes id_{\mathrm{A}})
\end{equation*}

2) a coassociatve comultiplication, $\Delta :\mathrm{A}\rightarrow \mathrm{A}%
\otimes \mathrm{A}$, $\ \mid \mathrm{a}\otimes \mathrm{b\mid =\mid a\mid }%
+\mid \mathrm{b\mid ,}$

$\Delta :\mathrm{A}_{\alpha }\subset \oplus _{\beta +\gamma =\alpha }\mathrm{%
A}_{\beta }\otimes \mathrm{A}_{\gamma },\Delta (\mathrm{a})=\sum_{i}\mathrm{a%
}_{i}^{(1)}\otimes \mathrm{b}_{i}^{(2)}$, \textrm{a}$\in $\textrm{A},%
\begin{equation*}
(id_{\mathrm{A}}\otimes \Delta )\circ \Delta =(\Delta \otimes id_{\mathrm{A}%
})\circ \Delta
\end{equation*}

3) a counit, $\varepsilon :\mathrm{A}\rightarrow \mathbf{C},$%
\begin{equation*}
(id_{\mathrm{A}}\otimes \varepsilon )\circ \Delta =(\varepsilon \otimes id_{%
\mathrm{A}})\circ \Delta =id_{\mathrm{A}}
\end{equation*}%
and we have $\varepsilon (\mathrm{A}_{1})=0$

4) an antipode $S:\mathrm{A}\rightarrow \mathrm{A,}$ $S$($\mathrm{A}_{\alpha
})\subset \mathrm{A}_{\alpha }$%
\begin{equation*}
m\circ (id_{\mathrm{A}}\otimes S)\circ \Delta =m\circ (S\otimes id_{\mathrm{A%
}})\circ \Delta =i\circ \varepsilon
\end{equation*}

such that the mappings $\Delta $ and $\varepsilon $ are algebra
homomorphisms $\mathtt{Z}_{2}$-graded algebras and in particular the
multiplication in $\mathrm{A}\otimes \mathrm{A}$ is given by
\begin{equation*}
(\mathrm{a}\otimes \mathrm{b})(\mathrm{c}\otimes \mathrm{d})=(-1)^{\mid
\mathrm{c}\mid \mid \mathrm{b\mid }}(\mathrm{ac}\otimes \mathrm{bd})
\end{equation*}
\end{definition}
%\end{definition}

One can show that the antipode $S$ is always an anti-homomorphism of the
algebra and of the coalgebra,%
\begin{equation*}
S(\mathrm{ab})=(-1)^{\mid \mathrm{a}\mid \mid \mathrm{b\mid }}S(\mathrm{a})S(%
\mathrm{b}),(S\otimes S)\circ \Delta =\tau \circ \Delta \circ S.
\end{equation*}%
where the map $\tau :\mathrm{A}\otimes \mathrm{A\rightarrow A}\otimes
\mathrm{A}$ is given by%
\begin{equation*}
\tau (\mathrm{a}\otimes \mathrm{b})=(-1)^{\mid \mathrm{a}\mid \mid \mathrm{%
b\mid }}\mathrm{b}\otimes \mathrm{a}
\end{equation*}%
We will need later on the following identity
\begin{equation}
\sum_{i,j}(\mathrm{a}_{i}^{(1)})_{j}^{(1)}\otimes S(\mathrm{a}%
_{i}^{(1)})_{j}^{(2)}\mathrm{a}_{i}^{(2)}=\mathrm{a}\otimes \mathbf{1}
\end{equation}%
where \textrm{a}$\in $\textrm{A. }This identity follows from coassociativity
of the coproduct $\Delta .$\qquad

The simplest example of $\mathtt{Z}_{2}$-graded Hopf algebra is the quantum
superalgebra $U_{q}[osp(1\mid 2)].$ The quantum superalgebra $%
U_{q}[osp(1\mid 2)]$ is $\mathtt{Z}_{2}$-graded algebra with unit $\mathbf{1}
$ and generated by three elements: $H$ ( $\deg (H)=0$) and $v_{\pm }$ ( $%
\deg (v_{\pm })=1$) with the follwing (anti)commutation relations%
\begin{equation}
\lbrack H,v_{\pm }]=\pm \frac{1}{2}v_{\pm };[v_{+},v_{-}]_{+}=-\frac{sh(\eta
H)}{sh(2\eta )}
\end{equation}%
where the parameter $\eta $ is real and we set $q=e^{-\frac{\eta }{2}}$. The
following formulae for coproduct $\Delta $, antipode $S$ and the counit $%
\varepsilon $ define on $U_{q}[osp(1\mid 2)]$ the structure of $\mathtt{Z}%
_{2}$-graded Hopf algebra
\begin{equation*}
\Delta (H)=H\otimes \mathbf{1+1\otimes }H;\Delta (v_{\pm })=v_{\pm }\otimes
q^{H}+q^{-H}\otimes v_{\pm },
\end{equation*}%
\begin{equation*}
\varepsilon (H)=\varepsilon (v_{\pm })=0,\varepsilon (\mathbf{1})=1
\end{equation*}%
and the antipode is defined by%
\begin{equation*}
S(H)=-H;S(v_{\pm })=-q^{\pm \frac{1}{2}}v_{\pm }.
\end{equation*}%
As of $\mathtt{Z}_{2}$-graded Hopf algebra $U_{q}[osp(1\mid 2)]$ has the
form $U_{q}[osp(1\mid 2)]=\oplus _{\alpha \in \mathtt{Z}_{2}}(U_{q}[osp(1%
\mid 2)])_{\alpha }$.

\bigskip For any $\mathtt{Z}_{2}$-graded Hopf algebra $\mathrm{A}$ one can
define the adjoint action $ad$ of $\mathrm{A}$ on itself in the following
way
\begin{equation*}
ad_{\mathrm{a}}(\mathrm{b})=\sum_{i}(-1)^{\mid \mathrm{a}_{i}^{(2)}\mid \mid
\mathrm{b\mid }}\mathrm{a}_{i}^{(1)}\mathrm{b}S(\mathrm{a}_{i}^{(2)})
\end{equation*}%
for any $\mathrm{a,}$ $\mathrm{b\in A}$. Using this action we define the
subset of invariant elements of $\mathrm{A}$
\begin{equation*}
\mathrm{A}_{\varepsilon }=\{\mathrm{b}\in \mathrm{A}:ad_{\mathrm{a}}(\mathrm{%
b})=\varepsilon (\mathrm{a)b},\forall \mathrm{a}\in \mathrm{A}\}.
\end{equation*}%
We will need later on the following proposition which characterises the
invariant elements of $\mathtt{Z}_{2}$-graded Hopf algebra $\mathrm{A}$

\begin{proposition}
An element $\mathrm{b\in }$ $\mathrm{A}$ is $ad$-invariant if and only if it
belongs to the center $Z(\mathrm{A})$ of $\mathrm{A}$ i.e. we have for any $%
\mathrm{a}$\ $\in $\textrm{A}
\begin{equation*}
ad_{\mathrm{a}}(\mathrm{b})=\sum_{i}(-1)^{\mid \mathrm{a}_{i}^{(2)}\mid \mid
\mathrm{b\mid }}(\mathrm{a}_{i}^{(1)})\mathrm{b}S(\mathrm{a}%
_{i}^{(2)}))=\varepsilon (\mathrm{a)b}\Leftrightarrow \mathrm{ab}=(-1)^{\mid
\mathrm{a}\mid \mid \mathrm{b\mid }}\mathrm{ba}
\end{equation*}%
or equivalently we have $\mathrm{A}_{\varepsilon }=Z(\mathrm{A})$.

\begin{proof}
First we prove \bigskip\ $(\Rightarrow )$. If $\mathrm{b\in }$ $Z(\mathrm{A)}
$ then we have for any \ $\mathrm{a}\in $\textrm{A}
\begin{equation*}
ad_{\mathrm{a}}(\mathrm{b})=\sum_{i}(-1)^{\mid \mathrm{a}_{i}^{(2)}\mid \mid
\mathrm{b\mid }}(\mathrm{a}_{i}^{(1)})\mathrm{b}S(\mathrm{a}%
_{i}^{(2)}))=\sum_{i}(\mathrm{a}_{i}^{(1)})(S(\mathrm{a}_{i}^{(2)}))\mathrm{%
b=}\varepsilon (\mathrm{a)b}
\end{equation*}%
The proof of the converse $(\Leftarrow )$ is more difficult. Now we assume
that $\mathrm{b\in }$ $\mathrm{A}_{\varepsilon }$ i.e. for any \ $\mathrm{a}%
\in $\textrm{A}
\begin{equation}
ad_{\mathrm{a}}(\mathrm{b})=\sum_{i}(-1)^{\mid \mathrm{a}_{i}^{(2)}\mid \mid
\mathrm{b\mid }}(\mathrm{a}_{i}^{(1)})\mathrm{b}S(\mathrm{a}%
_{i}^{(2)}))=\varepsilon (\mathrm{a)}f
\end{equation}%
and we have to prove that from this it follows%
\begin{equation}
\mathrm{ba=}(-1)^{\mid \mathrm{a}\mid \mid \mathrm{b\mid }}\mathrm{ab}
\end{equation}%
First let us observe that from $\varepsilon (\mathrm{A}_{1})=0$ we have for
any $\mathrm{a}$\ $\in $\textrm{A}%
\begin{equation}
\varepsilon (\mathrm{a)}=(-1)^{k\mid a\mathrm{\mid }}\varepsilon (\mathrm{a)}
\end{equation}%
where $k$ is arbitrary number. We have also from Definition 1 for any $i,j$
appearing in the coproduct $\Delta (\mathrm{a})$%
\begin{equation}
\mid \mathrm{a}_{i}^{(1)}\mid =\mid (\mathrm{a}_{i}^{(1)})_{j}^{(1)}\mid
+\mid \mathrm{a}_{i}^{(1)})_{j}^{(2)}\mid
\end{equation}%
We start from the LHS of the equation(2.4)%
\begin{equation*}
\mathrm{ba}=\sum_{i}\mathrm{b}[\varepsilon (\mathrm{a}_{i}^{(1)})\mathrm{a}%
_{i}^{(2)}]=\sum_{i}(-1)^{\mid \mathrm{a}_{i}^{(1)}\mid \mid \mathrm{b\mid }%
}\varepsilon (\mathrm{a}_{i}^{(1)})\mathrm{ba}_{i}^{(2)}
\end{equation*}%
where we have used the equation (2.5). Now we use the equation (2.3) for $%
\mathrm{a=a}_{i}^{(1)}$ and we get
\begin{equation*}
\mathrm{ba}=\sum_{ij}\{(-1)^{\mid \mathrm{a}_{i}^{(1)}\mid \mid \mathrm{%
b\mid }}(-1)^{\mid (\mathrm{a}_{i}^{(1)})_{j}^{(2)})\mid \mid \mathrm{b\mid }%
}[(\mathrm{a}_{i}^{(1)})_{j}^{(1)}]\mathrm{b}[S(\mathrm{a}%
_{i}^{(1)})_{j}^{(2)}]\}\mathrm{a}_{i}^{(2)}=
\end{equation*}%
\begin{equation*}
=\sum_{ij}(-1)^{\mid (\mathrm{a}_{i}^{(1)})_{j}^{(1)})\mid \mid \mathrm{%
b\mid }}[(\mathrm{a}_{i}^{(1)})_{j}^{(1)}]\mathrm{b}[S(\mathrm{a}%
_{i}^{(1)})_{j}^{(2)}\mathrm{a}_{i}^{(2)}].
\end{equation*}%
In the last equation we have used equation (2.6). Now we will prove that
\begin{equation*}
\sum_{ij}(-1)^{\mid (\mathrm{a}_{i}^{(1)})_{j}^{(1)})\mid \mid \mathrm{b\mid
}}[(\mathrm{a}_{i}^{(1)})_{j}^{(1)}]\mathrm{b}[S(\mathrm{a}%
_{i}^{(1)})_{j}^{(2)}\mathrm{a}_{i}^{(2)}]=(-1)^{\mid \mathrm{a}\mid \mid
\mathrm{b\mid }}\mathrm{ab}
\end{equation*}%
From the coassociativity condition for the coproduct $\Delta $ we get
\begin{equation*}
\sum_{i,j}\mathrm{b}\otimes (\mathrm{a}_{i}^{(1)})_{j}^{(1)}\otimes (\mathrm{%
a}_{i}^{(1)})_{j}^{(2)}\otimes \mathrm{a}_{i}^{(2)}=\sum_{i,j}\mathrm{b}%
\otimes \mathrm{a}_{i}^{(1)}\otimes (\mathrm{a}_{i}^{(2)})_{j}^{(1)}\otimes (%
\mathrm{a}_{i}^{(2)})_{j}^{(2)}
\end{equation*}%
Acting on both sides of the above equation by $(m\circ (m\otimes id)\circ
(m\otimes id\otimes id))\circ (\tau \otimes S\otimes id)$ we get
\begin{equation*}
\sum_{i,j}(-1)^{\mid (\mathrm{a}_{i}^{(1)})_{j}^{(1)})\mid \mid b\mathrm{%
\mid }}(\mathrm{a}_{i}^{(1)})_{j}^{(1)}\mathrm{b}S(\mathrm{a}%
_{i}^{(1)})_{j}^{(2)}\mathrm{a}_{i}^{(2)}=
\end{equation*}%
\begin{equation*}
=(-1)^{\mid \mathrm{a}\mid \mid \mathrm{b\mid }}\sum_{i,j}(-1)^{\mid \mathrm{%
a}_{i}^{(2)}\mid \mid b\mathrm{\mid }}\mathrm{a}_{i}^{(1)}\mathrm{b}S(%
\mathrm{a}_{i}^{(2)})_{j}^{(1)}(\mathrm{a}_{i}^{(2)})_{j}^{(2)}=
\end{equation*}%
\begin{equation*}
=(-1)^{\mid \mathrm{a}\mid \mid \mathrm{b\mid }}\sum_{i}(-1)^{\mid \mathrm{a}%
_{i}^{(2)}\mid \mid b\mathrm{\mid }}\mathrm{a}_{i}^{(1)}\mathrm{b}%
\varepsilon (\mathrm{a}_{i}^{(2)})=(-1)^{\mid \mathrm{a}\mid \mid \mathrm{%
b\mid }}\mathrm{ab}
\end{equation*}%
\qquad
\end{proof}
\end{proposition}

\bigskip In the following we will consider the representations of $\mathtt{Z}%
_{2}$-graded Hopf algebra $U_{q}[osp(1\mid 2)]$ in the $\mathtt{Z}_{2}$%
-graded linear spaces therefore we recall here some basic properties of the
graded representations \cite{17}. A vector space $V$\textrm{\ }over complex
field $\mathbf{C}$\ is called\textrm{\ }$\mathtt{Z}_{2}$-graded linear space
or simply graded space if $V$\textrm{\ }=$\oplus _{\alpha \in \mathtt{Z}%
_{2}}V_{\alpha }$\textrm{. }The elements $v$ of V$_{\alpha }$ are said to be
homogenous of degree $\alpha $ ($\alpha =0\leftrightarrow $even, $\alpha
=1\leftrightarrow $ odd) and their degree will be noted similarly as in case
of graded algebras $\deg (v)\equiv \mid v\mid \in \mathtt{Z}_{2}.$ Consider
now two graded vector spaces $V,W$ and a linear mapping $f\in Hom(V,W)$. The
mapping $f$ is said to be homogenous of degree $\beta \in \mathtt{Z}_{2}$ if
\begin{equation*}
f(V_{\alpha })\subset W_{\alpha +\beta }.
\end{equation*}%
where $\alpha \in \mathtt{Z}_{2}$ So we get a gradation in linear space $%
Hom(V,W)$
\begin{equation*}
Hom(V,W)_{\beta }=\{f\in Hom(V,W):f(V_{\alpha })\subset W_{\alpha +\beta }\}.
\end{equation*}%
and
\begin{equation*}
Hom(V,W)=Hom(V,W)_{0}\oplus Hom(V,W)_{1}
\end{equation*}

For a given $\mathtt{Z}_{2}$-graded Hopf algebra \textrm{A} a graded
representation of \textrm{A} is defined in the following way

\begin{definition}
A graded representation of $\mathtt{Z}_{2}$-graded Hopf algebra \textrm{A}
in $\mathtt{Z}_{2}$-graded linear space $V$\textrm{\ is }an even
homomorphism $\rho $ $:$ \textrm{A} $\rightarrow $ $Hom(V,V)$ i.e. $\rho \in
Hom($\textrm{A}$,$ $Hom(V,V))$. The pair $(V,\rho )$ is called a graded
representation of Hopf algebra \textrm{A}. The representation $(V,\rho )$ is
irreducible if there is no proper subspace $V^{^{\prime }}\subset V$ which
is invariant under action of the Hopf algebra \textrm{A} via map $\rho $.
\end{definition}

Let us recall some examples of $\mathtt{Z}_{2}$-graded Hopf algebra
representations.

\begin{example}
A $\mathtt{Z}_{2}$-graded Hopf algebra \textrm{A} is itself a graded
representation space for the adjoint action $\rho (\mathrm{a})\equiv ad_{%
\mathrm{a}}$
\begin{equation*}
ad_{\mathrm{a}}(\mathrm{b})=\sum_{i}(-1)^{\mid \mathrm{a}_{i}^{(2)}\mid \mid
\mathrm{b\mid }}\mathrm{a}_{i}^{(1)}\mathrm{b}S(\mathrm{a}_{i}^{(2)})
\end{equation*}

for $\mathrm{a},\mathrm{b}\in \mathrm{A.}$ This representation is denoted $($%
\textrm{A}$,ad)\equiv $\textrm{A}$_{ad}$.
\end{example}

\begin{example}
A $\mathtt{Z}_{2}$-graded Hopf algebra $\mathrm{A}$ is also a graded
representation space for a left regular action $L$ of \ $\mathrm{A}$
\begin{equation*}
L(\mathrm{a}).\mathrm{b}=m(\mathrm{a}\otimes \mathrm{b})=\mathrm{ab}
\end{equation*}%
for any $\mathrm{a},\mathrm{b}\in \mathrm{A.}$A left regular representation
is denoted $(\mathrm{A},L)\equiv \mathrm{A}_{L}.$
\end{example}

\begin{example}
Let $(V,\pi ),$ and $(W,\rho )$ be a graded modules of $\mathtt{Z}_{2}$%
-graded Hopf algebra $\mathrm{A}$. The linear space $Hom(V,W)$ is an graded
\textrm{A}-module $(Hom(V,W),\delta )$ with the action of \textrm{A} on$\
f\in Hom(V,W)$ defined as follows$\ \ \ \ \ \ \ \ \ \ \ \ \ \ \ \ \ \ \ \ \
\ \ \ \ \ \ \ $%
\begin{equation*}
\delta (\mathrm{a})(f)=\sum_{i}(-1)^{\mid \mathrm{a}_{i}^{(2)}\mid \mid f%
\mathrm{\mid }}\rho (\mathrm{a}_{i}^{(1)})\circ f\circ \pi (S(\mathrm{a}%
_{i}^{(2)})).
\end{equation*}
\end{example}

\begin{example}
The tensor product $V\otimes W$ of two graded representation spaces of
representations $(V,\pi ),$ and $(W,\rho )$ is a graded representation space
where the action $\delta ^{\otimes }$ of \textrm{A} is the followng \textrm{%
\ }%
\begin{equation*}
\delta ^{\otimes }(\mathrm{a)}(v\otimes w)=\sum_{i}(-1)^{\mid \mathrm{a}%
_{i}^{(2)}\mid \mid v\mathrm{\mid }}\pi (\mathrm{a}_{i}^{(1)})v\otimes \rho
(S(\mathrm{a}_{i}^{(2)}))w.
\end{equation*}%
for any $v\in V,w\in W$ and where $\mid v\otimes w\mathrm{\mid }=\mid v%
\mathrm{\mid +}\mid w\mathrm{\mid }$. This yields to the representation $%
(W\otimes V,(\rho \otimes \pi )\circ \Delta )$.
\end{example}

The last example is the following

\begin{example}
The counit map $\varepsilon $ of \ $\mathrm{A}$ equips any graded vector
space $V$ with a trivial representation $\rho =\varepsilon $ structure where
\begin{equation*}
\mathrm{a}v=\varepsilon (\mathrm{a)}v
\end{equation*}%
where $v\in V$ and $\mathrm{a}\in $\textrm{A. }In particular any
one-dimensional representation (wich is not a zero representation) is
equivalent to a trivial representation.
\end{example}

\bigskip The concept of trivial action of the $\mathtt{Z}_{2}$-graded Hopf
algebra $\mathrm{A}$ on vectors of representation space can be applied to
any representation of $\mathrm{A}$.

\begin{definition}
\bigskip For any representation $(V,\rho )$ of Hopf algebra $\mathrm{A}$ we
define the subspace of invariant vectors
\begin{equation*}
V_{\varepsilon }=\{v\in V:\rho (\mathrm{a}).v=\varepsilon (\mathrm{a)}%
v,\forall \mathrm{a}\in \mathrm{A}\}.
\end{equation*}
\end{definition}

\bigskip Next important matematical tool which we are going to use later on
is a graded intertwiner of representations so let us recall its definition.

\begin{definition}
Let $(V,\rho )$ and $(W,\sigma )$ be representations of the $\mathtt{Z}_{2}$%
-graded Hopf algebra $\mathrm{A}$. A linear map $f\in Hom(V,W)$ is a graded
intertwiner of representations $(V,\pi )$ and $(W,\rho )$ if%
\begin{equation*}
f\circ \pi (\mathrm{a})=(-1)^{\mid \mathrm{a}\mid \mid f\mathrm{\mid }}\rho (%
\mathrm{a})\circ f.
\end{equation*}%
for any \ $\mathrm{a}\in $\textrm{A. }The \ space of the graded intertwiners
will be denoted $I_{\mathrm{A}}(V,W)$. An even intertwiner is a homomorphism
of representations so the subspace ($I_{\mathrm{A}}(V,W))_{0}\equiv Hom_{%
\mathrm{A}}(V,W)$ is a space of homomorphisms.
\end{definition}

\bigskip We give two examples of homomorphisms of representations of \textrm{%
A}, which will be important in the following.

\begin{example}
\bigskip The $\mathtt{Z}_{2}$-graded Hopf algebra $\mathrm{A}$ with the
adjoint action $ad_{\mathrm{a}}$, $\mathrm{a}\in \mathrm{A}$ form the
adjoint representation $(\mathrm{A},ad_{\mathrm{a}})$. On the other hand we
have the representation $(Hom(V,V),\delta )$ of $\mathrm{A}$ from Example 3.
The representation $\rho $ $:$ \textrm{A} $\rightarrow $ $Hom(V,V)$ from
Definition 2 is a homomorphism of the Hopf algebra representations.
\end{example}

\begin{example}
A left regular action $L$ given in Example 2 is a homomorphism of
representations $\mathrm{A}_{ad}$ and ($Hom(\mathrm{A}_{L},\mathrm{A}%
_{L}),\delta )$ i.e. $L\in Hom_{\mathrm{A}}($ $\mathrm{A}_{ad},Hom(\mathrm{A}%
_{L},\mathrm{A}_{L})).$ In fact we have for any $\mathrm{a},\mathrm{b}\in
\mathrm{A}$
\begin{equation*}
L(ad_{\mathrm{a}}(\mathrm{b}))=\sum_{i}(-1)^{\mid \mathrm{a}_{i}^{(2)}%
\mathrm{\mid \mid b}\mathit{\mid }}L(\mathrm{a}_{i}^{(1)}\mathrm{b}S(\mathrm{%
a}_{i}^{(2)}))=\sum_{i}(-1)^{\mid \mathrm{a}_{i}^{(2)}\mathrm{\mid \mid b}%
\mathit{\mid }}L(\mathrm{a}_{i}^{(1)})L(\mathrm{b)}L(S(\mathrm{a}_{i}^{(2)}))
\end{equation*}
or equivalently
\begin{equation*}
L\circ ad_{\mathrm{a}}=\delta (\mathrm{a})\circ L.
\end{equation*}%
where $\mid L\mid =0$ because $L$ is a representation.
\end{example}

\bigskip Now we are in the position to define tensor operators for $\mathtt{Z%
}_{2}$-graded Hopf algebras. \bigskip Following the idea of the definition
of tensor operators for Hopf algebras given in \cite{12} we define tensor
operators for $\mathtt{Z}_{2}$-graded Hopf algebras in the following way

\begin{definition}
Let $(V,\pi )$, $(W,\rho )$ and $(U,\sigma )$ be graded representations of
the $\mathtt{Z}_{2}$-graded Hopf algebra $\mathrm{A}$ and let $\mathit{T}\in
Hom(V,Hom(W,U))$ then $\mathit{T}$ is a tensor operator of type $V$ in $W$
if $\mathit{T}\in I_{\mathrm{A}}(V,Hom(W,U))$ . In other words tensor
operator $\mathit{T}$ is a graded intertwiner of \ representations\textrm{\ }
$(V,\pi )$ and $(Hom(W,U),\delta )$ and it satisfies
\begin{equation}
\mathit{T}\circ \pi (\mathrm{a})=(-1)^{\mid \mathrm{a\mid \mid }\mathit{%
T\mid }}\delta (\mathrm{a})\circ \mathit{T.}
\end{equation}

Let vectors $\{e_{l}\}_{l\in I\subset N}$ be a basis of the representation
space $V$, then the linear operators $\mathit{T}(e_{l})\equiv \mathit{T}%
_{l}\in Hom(W,U)$ will be called the components of the tensor operator $%
\mathit{T}$. If $\dim V<\infty $ then the components $\mathit{T}_{l}$ of $%
\mathit{T}$ satisfie%
\begin{equation}
\pi (\mathrm{a})_{jl}\mathit{T}_{j}=(-1)^{\mid \mathrm{a\mid \mid }\mathit{%
T\mid }}\sum_{i}(-1)^{\mid \mathrm{a}_{i}^{(2)}\mathrm{\mid \mid }\mathit{T}%
_{l}\mathit{\mid }}\sigma (\mathrm{a}_{i}^{(1)})\circ \mathit{T}_{l}\circ
\rho (S(\mathrm{a}_{i}^{(2)}))
\end{equation}%
where $\pi (\mathrm{a})_{jl}$ is a matrix of $\pi (\mathrm{a})$. If all the
representations $(V,\pi )$, $(W,\rho )$ and $(U,\sigma )$ are irreducibles
then the tensor operator $\mathit{T}$ is called irreducible.
\end{definition}

\bigskip \bigskip\ Let write the defining equation (2.8) for the components $%
\mathit{T}_{l}$ of $\mathit{T}$ when $\mathrm{A=}$ $U_{q}[osp(1\mid 2)]$ and
$\mathrm{a}=v_{\pm },H$
\begin{equation}
\pi (v_{+})_{jl}\mathit{T}_{j}=(-1)^{\mid v_{+}\mathrm{\mid \mid }\mathit{%
T\mid }}(\sigma (v_{+})\circ \mathit{T}_{l}\circ \rho (q^{-H})-(-1)^{\mid
v_{+}\mathrm{\mid \mid }\mathit{T}_{l}\mathit{\mid }}q^{\frac{1}{2}}\sigma
(q^{-H})\circ \mathit{T}_{l}\circ \rho (v_{+}))
\end{equation}%
\begin{equation}
\pi (v_{-})_{jl}\mathit{T}_{j}=(-1)^{\mid v_{-}\mathrm{\mid \mid }\mathit{%
T\mid }}(\sigma (v_{-})\circ \mathit{T}_{l}\circ \rho (q^{-H})-(-1)^{\mid
v_{-}\mathrm{\mid \mid }\mathit{T}_{l}\mathit{\mid }}q^{-\frac{1}{2}}\sigma
(q^{-H})\circ \mathit{T}_{l}\circ \rho (v_{-}))
\end{equation}%
\begin{equation}
\pi (H)_{jl}\mathit{T}_{j}=\sigma (H)\circ \mathit{T}_{l}-\mathit{T}%
_{l}\circ \rho (H)
\end{equation}

Thus the above definition of tensor operator although seems to be abstract
in case of the simplest quantum superalgebra $U_{q}[osp(1\mid 2)]$ which is
a superanalogue of the quantum algebra $U_{q}[su(2))],$ gives very similar
defining formulae for generating elements as in the case of $U_{q}[su(2))]$
\cite{6, 7, 10}.

Let us give some important example of tensor operator.

\begin{example}
The Example 6 shows that the representatoin $\rho $ from Definition 2 is
itself a tensor operator because $\rho \in Hom_{\mathrm{A}}(\mathrm{A}%
,Hom(W\otimes W)).$
\end{example}

\begin{example}
The left regular action $L$ of $\mathrm{A}$ on itself as defined in Example
2 is a tensor operator because $L\in Hom_{\mathrm{A}}($ $\mathrm{A}_{ad},Hom(%
\mathrm{A}_{L},\mathrm{A}_{L}))$ (Example 7).
\end{example}

Before formulation a lemma which will be used later on we introduce a useful
notation If $f\in Hom(V,W)$ where $(V,\pi )$ and $(W,\rho )$ are
representations of the Hopf algebra $\mathrm{A}$ then we define%
\begin{equation}
\pi _{f}(\mathrm{a})\equiv f\circ \pi (\mathrm{a}):V\rightarrow W
\end{equation}%
and the linear mapping $m_{\rho }^{\pi }:Hom(W)\otimes Hom(V,W)\rightarrow
Hom(V,W)$ is defined in the following way%
\begin{equation}
m_{\rho }^{\pi }(\rho (\mathrm{a})\otimes \pi _{f}(\mathrm{b}))=(-1)^{\mid
\mathrm{a\mid \mid }\mathit{f\mid }}((m_{\rho }^{\pi }\circ (\rho \otimes
\pi _{f})).(\mathrm{a}\otimes \mathrm{b})\equiv \rho (\mathrm{a})\circ \pi
_{f}(\mathrm{b}).
\end{equation}

\begin{lemma}
Assume that

1) $(V,\pi ),W,\rho ),$ $(U,\sigma )$ and $(Hom(W,U),\delta )$ are
representations of the $\mathtt{Z}_{2}$-graded Hopf algebra $\mathrm{A,}$

2) $\mathit{T}\in I_{\mathrm{A}}(V,Hom(W,U))$\ \ \ \ i.e. $\forall \mathrm{%
a\in A}$ $\mathit{T}\circ \pi (\mathrm{a})=(-1)^{\mid \mathrm{a\mid \mid }%
\mathit{T\mid }}\delta (\mathrm{a})\circ \mathit{T}$ $,$

3) $\mathit{\check{T}}\in Hom(V\otimes W,U))$ and $\check{T}(v\otimes
w)\equiv \mathit{T}(v).w$ \ $\forall v\in V,w\in W.$ \

\ Then \ \ \ \ \ \ \ \ \ \ \ \ \ \ \ \ \ \ \ \ \ \ \ \ \ \ \ \ \ \ \ \ \ \ \
\ \ \ \ \ \ \ \ \ \ \ \ \ \ \ \ \ \ \ \ \ \ \ \ \ \ \ \ \ \ \ \ \ \ \ \ \

a) $\mathit{\check{T}}\in I_{\mathrm{A}}(V\otimes W,U))$ \ i.e. $\forall
\mathrm{a\in A}$ $\mathit{\check{T}}\circ \lbrack (\pi \otimes \rho )\Delta (%
\mathrm{a})]=(-1)^{\mid \mathrm{a\mid \mid }\mathit{\check{T}\mid }}\sigma (%
\mathrm{a})\circ \mathit{\check{T}}$.

b) $\mathrm{\mid }\mathit{T\mid =}\mathrm{\mid }\mathit{\check{T}\mid }$
\end{lemma}

\begin{proof}
Let us prove $a).$ The action $\delta $ of representation $(Hom(W,U),\delta
) $ is given in Example 3. We rewrite the condition 2) for $\mathit{T}$ in
the form

\begin{equation}
\mathit{T}[\pi (\mathrm{a}).v].w=(-1)^{\mid \mathrm{a\mid \mid }\mathit{%
T\mid }}\{\sum_{i}(-1)^{\mid \mathrm{a}_{i}^{(2)}\mid \mid \mathit{T(}v)%
\mathrm{\mid }}\sigma (\mathrm{a}_{i}^{(1)})\circ \mathit{T}(v)\circ \rho (S(%
\mathrm{a}_{i}^{(2)}))\}.w
\end{equation}%
for any $\mathrm{a\in A}$, $v\in V$, $w\in W.$ We have to prove that from
this it follows condition a) for $\mathit{\check{T}}$ which can be written
as follows%
\begin{equation}
\sum_{i}(-1)^{\mid \mathrm{a}_{i}^{(2)}\mid \mid v\mathrm{\mid }}\mathit{T}%
[\pi (\mathrm{a}_{i}^{1}).v].(\rho (\mathrm{a}_{i}^{(2)}).w)=(-1)^{\mid
\mathrm{a\mid \mid }\mathit{T\mid }}\sigma (\mathrm{a})[\mathit{T}(v).w]
\end{equation}%
for any $\mathrm{a\in A}$, $v\in V$, $w\in W.$ Applying in LHS of the above
equation condition (2.14 ) for $\mathrm{a}=\mathrm{a}_{i}^{1}$ \ we get
\begin{equation*}
\sum_{i}(-1)^{\mid \mathrm{a}_{i}^{(2)}\mid \mid v\mathrm{\mid }}\mathit{T}%
[\pi (\mathrm{a}_{i}^{1}).v].(\rho (\mathrm{a}_{i}^{(2)}).w)=\sum_{ij}(-1)^{%
\mid \mathrm{a}_{i}^{(2)}\mid \mid v\mathrm{\mid }}(-1)^{\mid \mathrm{a}%
_{i}^{1}\mathrm{\mid \mid }\mathit{T\mid }}\times
\end{equation*}%
\begin{equation*}
\times (-1)^{\mid (\mathrm{a}_{i}^{(1)})_{j}^{(2)}\mid \mid \mathit{T(}v)%
\mathrm{\mid }}\sigma \lbrack (\mathrm{a}_{i}^{(1)})_{j}^{(1)}]\circ \mathit{%
T}(v)\circ \rho \lbrack S(\mathrm{a}_{i}^{(1)})_{j}^{(2)}(\mathrm{a}%
_{i}^{(2)})].w
\end{equation*}%
In the notation (2.12), (2.13) it takes the form
\begin{equation*}
\sum_{i}(-1)^{\mid \mathrm{a}_{i}^{(2)}\mid \mid v\mathrm{\mid }}\mathit{T}%
[\pi (\mathrm{a}_{i}^{1}).v].(\rho (\mathrm{a}_{i}^{(2)}).w)=\sum_{ij}(-1)^{%
\mid \mathrm{a}_{i}^{(2)}\mid \mid v\mathrm{\mid }}(-1)^{\mid \mathrm{a}%
_{i}^{1}\mathrm{\mid \mid }\mathit{T\mid }}(-1)^{\mid (\mathrm{a}%
_{i}^{(1)})_{j}^{(2)}\mid \mid \mathit{T(}v)\mathrm{\mid }}\times
\end{equation*}%
\begin{equation*}
\times (-1)^{\mid (\mathrm{a}_{i}^{(1)})_{j}^{(1)}\mid \mid \mathit{T(}v)%
\mathrm{\mid }}\{m_{\sigma }^{\rho }\circ (\sigma \otimes \rho _{\mathit{T}%
(v)}).((\mathrm{a}_{i}^{(1)})_{j}^{(1)}\otimes S(\mathrm{a}%
_{i}^{(1)})_{j}^{(2)}\mathrm{a}_{i}^{(2)})\}.w
\end{equation*}%
Simpifying the phase and using the identity (2.1) we get
\begin{eqnarray*}
\sum_{i}(-1)^{\mid \mathrm{a}_{i}^{(2)}\mid \mid v\mathrm{\mid }}\mathit{T}%
[\pi (\mathrm{a}_{i}^{1}).v].(\rho (\mathrm{a}_{i}^{(2)}).w) &=&(-1)^{\mid
\mathrm{a\mid \mid }v\mathit{\mid }}\{m_{\sigma }^{\rho }\circ (\sigma
\otimes \rho _{\mathit{T}(v)}).(\mathrm{a}\otimes \mathbf{1})\}.w \\
&=&(-1)^{\mid \mathrm{a\mid \mid }v\mathit{\mid }}(-1)^{\mid \mathrm{a\mid
\mid }\mathit{T}(v)\mathit{\mid }}\sigma (\mathrm{a})[\mathit{T}(v).w] \\
&=&(-1)^{\mid \mathrm{a\mid \mid }\mathit{v\mid }+\mid \mathrm{a\mid \mid }%
\mathit{T}(v)\mathit{\mid }}\sigma (\mathrm{a})[\mathit{T}(v).w]
\end{eqnarray*}%
Which is RHS of \ the equation (2.15). The statement b) can be proved
considering the degrees of the values of $\mathit{T}$ and $\mathit{\check{T}}
$ on homogenous arguments.
\end{proof}

\section{\protect\bigskip Wigner-Eckart theorem for the quantum superalgebra
$U_{q}[osp(1\mid 2)]$}

In this section we will consider the quantum superalgebra $U_{q}[osp(1\mid
2)]$ and its graded representations. \bigskip A representation of \ the
quantum superalgebra $U_{q}[osp(1\mid 2)]$ in the graded linear space $V$
will be denoted by $\pi $
\begin{equation*}
\pi :U_{q}[osp(1\mid 2)]\rightarrow Hom(V,V).
\end{equation*}%
The finite dimensional irreducible representations of $U_{q}[osp(1\mid 2)]$
\ has been studied firstly in \cite{15}. They have the same structure as in
case of the nondefrmed superalgebra $osp(1\mid 2)$ and for this superalgebra
every finite dimensional irreducible representation is equivalent to a grade
star representation \cite{16}. It has been shown in \cite{2} that any finite
dimensional grade star representation of $U_{q}[osp(1\mid 2)]$ is
characterized by four parameters: the highest wieght $l$ (a non-negative
integer), the parity $\lambda =0,1$ of the highest wieght vector in the
representation space and by $\varphi ,\psi =0,1$, the signature parameters
of the Hermitean in the representation space $V.$ The parity $\lambda $ and
the signature $\varphi $ define the class $\epsilon =0,1$ of the grade star
representation by
\begin{equation*}
%\epsilon =\lambda +\varphi +1,\func{\rm mod}(2).
\epsilon =\lambda+\varphi +1,{\rm mod}(2).
\end{equation*}%
For simplicity we will write $(V^{l}(\lambda ),\pi ^{l})$ instead $%
(V^{l}(\lambda ),\pi _{\varphi \psi }^{l\epsilon })$ The representation
space $V^{l}(\lambda )$ is a graded vector space of dimension $2l+1$ with
basis $e_{m}^{l}(\lambda )$ where $-l\leq m\leq l$. The parity of the basis
vectors $e_{m}^{l}(\lambda )$ in determined by values of $l,m$ and $\lambda $%
\begin{equation*}
\mid e_{m}^{l}(\lambda )\mid =l-m+\lambda {\rm mod}(2).
\end{equation*}%
The vectors $e_{m}^{l}(\lambda )$ are pseudo-orthogonal with respect to the
Hermitean foem in $V$ and their normalisation is determined by the signature
parameters $\varphi ,\psi $
\begin{equation*}
(e_{m}^{l}(\lambda ),e_{m^{^{\prime }}}^{l^{^{\prime }}}(\lambda
))=(-1)^{\varphi (l-m)+\psi }\delta _{mm^{^{\prime }}},
\end{equation*}%
where $(,)$ denotes the Hermitean form in the representation space $%
V^{l}(\lambda )$. The operatorts $\pi ^{l}(v_{\pm })$ and $\pi ^{l}(H)$ act
on the basis $e_{m}^{l}(\lambda )$ in the following way%
\begin{equation}
\pi ^{l}(v_{+}).e_{m}^{l}=(-1)^{(l-m)}([l-m][l+m+1]\gamma )^{\frac{1}{2}%
}e_{m+1}^{l}
\end{equation}%
\begin{equation}
\pi ^{l}(v_{-}).e_{m}^{l}=([l+m][l-m+1]\gamma )^{\frac{1}{2}}e_{m-1}^{l}
\end{equation}%
\begin{equation}
\pi ^{l}(H).e_{m}^{l}=\frac{m}{2}e_{m}^{l}
\end{equation}%
where $[n]=\frac{q^{-\frac{n}{2}}-(-1)^{n}q^{\frac{n}{2}}}{q^{-\frac{1}{2}%
}-q^{\frac{1}{2}}}.$Note that the action of the operators $\pi ^{l}(v_{\pm
}) $ and $\pi ^{l}(H)$ does not depend on the parameters $\lambda ,\varphi
,\psi .$

Tensor product of two irreducible representation $(V^{l_{1}}(\lambda
_{1}),\pi ^{l_{1}})$ and $(V^{l_{2}}(\lambda _{2}),\pi ^{l_{2}})$ is
completely and simply reducible i.e. we have
\begin{equation*}
V^{l_{1}}(\lambda _{1})\otimes V^{l_{2}}(\lambda _{2})=\oplus _{l=\mid
l_{1}-l_{2}\mid }^{l_{1}+l_{2}}V^{l}(\lambda ).
\end{equation*}%
By definition the Glebsch-Gordan coefficients (C-Gc) $(l_{1}m_{1}\lambda
_{1},l_{2}m_{2}\lambda _{2}\mid lm\lambda )_{q}$ relate the standard basis \
$e_{m_{1}}^{l_{1}}(\lambda _{1})\otimes e_{m_{2}}^{l_{2}}(\lambda _{2})$ of
tensor product $V^{l_{1}}(\lambda _{1})\otimes V^{l_{2}}(\lambda _{2})$ to
the reduced basis $e_{m}^{l}(l_{1},l_{2},\lambda )$ in the following way
\begin{equation*}
e_{m}^{l}(l_{1},l_{2},\lambda )=\sum_{m_{1}m_{2}}(l_{1}m_{1}\lambda
_{1},l_{2}m_{2}\lambda _{2}\mid lm\lambda )_{q}e_{m_{1}}^{l_{1}}(\lambda
_{1})\otimes e_{m_{2}}^{l_{2}}(\lambda _{2})
\end{equation*}%
or equivalently%
\begin{equation*}
(-1)^{(l_{1-}m_{1})(l_{2-}m_{2})}e_{m_{1}}^{l_{1}}(\lambda _{1})\otimes
e_{m_{2}}^{l_{2}}(\lambda _{2})=
\end{equation*}%
\begin{equation*}
=\sum_{lm}(-1)^{(l-m)L}(l_{1}m_{1}\lambda _{1},l_{2}m_{2}\lambda _{2}\mid
lm\lambda )_{q}e_{m}^{l}(l_{1},l_{2},\lambda )
\end{equation*}%
where $m_{1}+m_{2}=m,L=l_{1}+l_{2}+l$ and $l$ is an integer satisfying the
condition
\begin{equation*}
\mid l_{1}-l_{2}\mid \leq l\leq l_{1}+l_{2}.
\end{equation*}%
In the following, in order to get Wigner-Eckart theorem in a conventional
form we will use a modified C-Gc $[l_{1}m_{1}\lambda _{1},l_{2}m_{2}\lambda
_{2}\mid lm\lambda ]_{q}$ which are related to $(l_{1}m_{1}\lambda
_{1},l_{2}m_{2}\lambda _{2}\mid lm\lambda )_{q}$ by
\begin{equation*}
\lbrack l_{1}m_{1}\lambda _{1},l_{2}m_{2}\lambda _{2}\mid lm\lambda
]_{q}=(-1)^{(l_{1-}m_{1})(l_{2-}m_{2})}(-1)^{(l-m)L}(l_{1}m_{1}\lambda
_{1},l_{2}m_{2}\lambda _{2}\mid lm\lambda )_{q}.
\end{equation*}%
In terms of the modified C-Gc the relation between standard and reduced
basis in $V^{l_{1}}(\lambda _{1})\otimes V^{l_{2}}(\lambda _{2})$ looks%
\begin{equation*}
(-1)^{(l-m)L}e_{m}^{l}(l_{1},l_{2},\lambda )=
\end{equation*}%
\begin{equation*}
=\sum_{m_{1}m_{2}}(-1)^{(l_{1-}m_{1})(l_{2-}m_{2})}[l_{1}m_{1}\lambda
_{1},l_{2}m_{2}\lambda _{2}\mid lm\lambda ]_{q}e_{m_{1}}^{l_{1}}(\lambda
_{1})\otimes e_{m_{2}}^{l_{2}}(\lambda _{2})
\end{equation*}%
or equivalently
\begin{equation}
e_{m_{1}}^{l_{1}}(\lambda _{1})\otimes e_{m_{2}}^{l_{2}}(\lambda
_{2})=\sum_{lm}[l_{1}m_{1}\lambda _{1},l_{2}m_{2}\lambda _{2}\mid lm\lambda
]_{q}e_{m}^{l}(l_{1},l_{2},\lambda ).
\end{equation}%
We have also for any $l,m$ in this decoposition
\begin{equation}
\mid e_{m_{1}}^{l_{1}}(\lambda _{1})\otimes e_{m_{2}}^{l_{2}}(\lambda
_{2})\mid =\mid e_{m}^{l}(l_{1},l_{2},\lambda )\mid
\end{equation}%
In the classical theory of Racah-Wigner calculus, a very important role is
played by the C-Gc $(jm,jn\mid 00)$, which defines an invarint metric. In
the case of the quantum superalgebra $U_{q}[osp(1\mid 2)],$ the
corresponding coefficient also defines an invariant metric. It has the form%
\begin{equation}
C_{mn}^{l}(\lambda )=\sqrt{[2l+1]}(lm\lambda ,\ln \lambda \mid
00)_{q}=(-1)^{(l-m)\lambda }(-1)^{(l-m)(l-m-1)/2}q^{m/2}\delta m,-n.
\end{equation}%
For more details on the irreducible grade star representations and
properties of C-Gc see \cite{2}.

In case of \ the irreducible finite dimensional representations of the
quantum superalgebra $U_{q}[osp(1\mid 2)]$ Schur lemma has the following form

\begin{lemma}
\bigskip Let $(V^{l_{1}}(\lambda _{1}),\pi ^{l_{1}})$ and $%
(V^{l_{2}}(\lambda _{2}),\pi ^{l_{2}})$ be irreducible finite dimensional
representations of $U_{q}[osp(1\mid 2)]$ and let \bigskip $f\in
I_{U_{q}[osp(1\mid 2)]}(V^{l_{1}}(\lambda _{1}),V^{l_{2}}(\lambda _{2}))$
i.e. for any $a\in U_{q}[osp(1\mid 2)]$, $x\in V^{l_{1}}(\lambda _{1})$%
\begin{equation}
f(\pi ^{l_{1}}(a).x)=(-1)^{\mid f\mid \mid a\mid }\pi ^{l_{2}}(a)f(x),
\end{equation}%
then $f=\alpha id_{V^{l_{1}}(\lambda _{1})}$ $(\alpha \in \mathrm{R})$ if $\
l_{1}=l_{2}$ and $\lambda _{1}=\lambda _{2}$, or $f=0$ if $l_{1}\neq l_{2}$
or $\lambda _{1}\neq \lambda _{2}.$
\end{lemma}

\begin{proof}
\bigskip Let us consider the properties of the vector%
\begin{equation*}
y=f(e_{l_{1}}^{l_{1}}(\lambda _{1}))\in V^{l_{2}}(\lambda _{2}).
\end{equation*}%
Using equation (3.7) we get
\begin{equation*}
\pi ^{l_{2}}(H).y=\frac{l_{1}}{2}y;\pi ^{l_{2}}(v_{+}).y=0
\end{equation*}%
so either $y\in $ $V^{l_{2}}(\lambda _{2})$ is the highest weight vector of
weight $l_{1}$ in $V^{l_{2}}(\lambda _{2})$ or $f=0$ i.e. either $l_{1}=l_{2}
$ or $f=0.$ Assume that $l_{1}=l_{2}$ and $\lambda _{1},\lambda _{2}$
arbitrary. Then from the above it follows that we have
\begin{equation}
f(e_{m_{1}}^{l_{1}}(\lambda _{1}))=\alpha e_{m_{1}}^{l_{1}}(\lambda _{2})
\end{equation}%
and $\mid f\mid =1$ if $\lambda _{1}+\lambda _{2}=1$ or $\mid f\mid =0$ if $%
\lambda _{1}+\lambda _{2}=0$ mod(2). Acting on both sides of the
above equation by $T^{l_{1}}(v_{+})$ we get
\begin{equation*}
(-1)^{(l_{1}-m_{1})}([l_{1}-m_{1}][l_{1}+m_{1}+1]\gamma )^{\frac{1}{2}%
}e_{m_{1}+1}^{l}(\lambda _{2})=
\end{equation*}%
\begin{equation*}
=(-1)^{\mid f\mid }(-1)^{(l_{1}-m_{1})}([l_{1}-m_{1}][l_{1}+m_{1}+1]\gamma
)^{\frac{1}{2}}e_{m_{1}+1}^{l}(\lambda _{2})
\end{equation*}%
so  $f=0$ \ if $\ \lambda _{1}+\lambda _{2}=1.$
\end{proof}

We will need later on the following proposition which is a consequence of
Schur lemma

\begin{proposition}
\bigskip Let $(V^{l_{1}}(\lambda _{1}),\pi ^{l_{1}}),$ $(V^{l_{2}}(\lambda
_{2}),\pi ^{l_{2}})$ and $(V^{l_{3}}(\lambda _{3}),\pi ^{l_{3}})$ be
irreducible finite dimensional representations of $U_{q}[osp(1\mid 2)]$ with
bases respectively $\{e_{m_{1}}^{l_{2}}(\lambda _{1})\},$ $%
\{e_{m_{2}}^{l_{1}}(\lambda _{2})\}$, $\{e_{m_{3}}^{l_{3}}(\lambda _{3})\}$
and let \bigskip $f\in I_{U_{q}[osp(1\mid 2)]}(V^{l_{1}}(\lambda
_{1})\otimes V^{l_{2}}(\lambda _{2})),V^{l_{3}}(\lambda _{3}))$ where $\mid
l_{1}-l_{2}\mid \leq l_{3}\leq l_{1}+l_{2}.$ Then
\begin{equation}
f(e_{m_{1}}^{l_{1}}(\lambda _{1})\otimes e_{m_{2}}^{l_{2}}(\lambda
_{2}))=\alpha _{l_{3}}\sum_{m_{3}}[l_{1}m_{1}\lambda _{1},l_{2}m_{2}\lambda
_{2}\mid l_{3}m_{3}\lambda _{3}]_{q}e_{m_{3}}^{l_{3}}(l_{1},l_{2},\lambda
_{3}).
\end{equation}%
for any $e_{m_{i}}^{l_{i}}(\lambda _{i})\in V^{l_{i}}(\lambda _{i}),$ $i=1,2$
and $f$ $\in (I_{U_{q}[osp(1\mid 2)]}(V^{l_{1}}(\lambda _{1})\otimes
V^{l_{2}}(\lambda _{2})),V^{l_{3}}(\lambda _{3}))_{0}$ i.e. $f$ is an
homomorphism.
\end{proposition}

\begin{proof}
From Clebsch-Gordan decomposition we have
\begin{equation}
e_{m_{1}}^{l_{1}}(\lambda _{1})\otimes e_{m_{2}}^{l_{2}}(\lambda
_{2})=\sum_{lm}[l_{1}m_{1}\lambda _{1},l_{2}m_{2}\lambda _{2}\mid lm\lambda
]_{q}e_{m}^{l}(l_{1},l_{2},\lambda )
\end{equation}%
and for any $\mid l_{1}-l_{2}\mid \leq l\leq l_{1}+l_{2}$ the linear mapping
$f_{l}=f\mid _{V^{l}(\lambda )}:V^{l}(\lambda )\rightarrow V^{l_{3}}(\lambda
_{3})$ is an intertwiner of representations $V^{l}(\lambda )$ and $%
V^{l_{3}}(\lambda _{3})$ i.e. $f_{l}$ $\in I_{U_{q}[osp(1\mid
2)]}(V^{l}(\lambda ),V^{l_{3}}(\lambda _{3})).$ Therefore we have from Schur
lemma
\begin{equation*}
f_{l}=\alpha _{l}id_{V^{l}(\lambda )}\delta _{ll_{3}}\delta _{\lambda
\lambda _{3}}
\end{equation*}%
Taking into account that $f=\oplus _{l}f_{l}$ we get from the Clebsch-Gordan
decomposition (3.10) the equation (3.9) and it is clear that $\alpha _{l}$
do not depend on $m_{1}m_{2},m$. The fact that $f$ $\in (I_{U_{q}[osp(1\mid
2)]}(V^{l_{1}}(\lambda _{1})\otimes V^{l_{2}}(\lambda
_{2})),V^{l_{3}}(\lambda _{3}))_{0}$ follows from the relation (3.5).
\end{proof}

Now we can formulate Wigner-Eckart theorem for irreducible tensor operators
the for quantum superalgebra $U_{q}[osp(1\mid 2)]$.

\begin{theorem}
If $\mathit{T}\in I_{U_{q}[osp(1\mid 2)]}(V^{l_{1}}(\lambda
_{1}),Hom(V^{l_{2}}(\lambda _{2}),V^{l_{3}}(\lambda _{3})))$ is an
irreducible tensor operator .Then

1) the matrix elements of its components $\mathit{T}(e_{m_{1}}^{l_{1}}(%
\lambda _{1}))$ are proprtional to the modified Clebsch-Gordan coefficients
i.e.%
\begin{equation*}
\lbrack \mathit{T}(e_{m_{1}}^{l_{1}}(\lambda _{1}))]_{m_{3}m_{2}}=\alpha
\lbrack l_{1}m_{1}\lambda _{1},l_{2}m_{2}\lambda _{2}\mid l_{3}m_{3}\lambda
_{3}]_{q}
\end{equation*}%
where $\alpha $ is a real number called reduced matrix element which do not
depend on $m_{i},i=1,2,3.$

2) $\mathit{T}$ is an even intertwiner i.e. $\mathit{T}\in
Hom_{U_{q}[osp(1\mid 2)]}(V^{l_{1}}(\lambda _{1}),Hom(V^{l_{2}}(\lambda
_{2}),V^{l_{3}}(\lambda _{3})))$
\end{theorem}

\begin{proof}
From Lemma 1 we know that linear mapping $\mathit{\check{T}}\in
Hom(V^{l_{1}}(\lambda _{1})\otimes V^{l_{2}}(\lambda _{2}),V^{l_{3}}(\lambda
_{3})))$, $\mid l_{1}-l_{2}\mid \leq l_{3}\leq l_{1}+l_{2}$
\begin{equation*}
\mathit{\check{T}}(e_{m_{1}}^{l_{1}}(\lambda _{1})\otimes
e_{m_{2}}^{l_{2}}(\lambda _{2}))=\mathit{T}(e_{m_{1}}^{l_{1}}(\lambda
_{1})).e_{m_{2}}^{l_{2}}(\lambda _{2})
\end{equation*}%
is an intertwiner of representations and $\mathrm{\mid }\mathit{T\mid =}%
\mathrm{\mid }\mathit{\check{T}\mid }$. Then from Proposition 2 we get%
\begin{eqnarray*}
\mathit{\check{T}}(e_{m_{1}}^{l_{1}}(\lambda _{1})\otimes
e_{m_{2}}^{l_{2}}(\lambda _{2})) &=&\mathit{T}(e_{m_{1}}^{l_{1}}(\lambda
_{1})).e_{m_{2}}^{l_{2}}(\lambda _{2})= \\
&=&\alpha \sum_{m_{3}}[l_{1}m_{1}\lambda _{1},l_{2}m_{2}\lambda _{2}\mid
l_{3}m_{3}\lambda _{3}]_{q}e_{m_{3}}^{l_{3}}(l_{1},l_{2},\lambda _{3})
\end{eqnarray*}%
where $\alpha $ do not depend on $m_{i},i=1,2,3$ and $\mathit{\check{T}}$ is
even$.$On the other hand the matrix of the operator $\mathit{T}%
(e_{m_{1}}^{l_{1}}(\lambda _{1}))$ is defined by equation%
\begin{equation*}
\mathit{T}(e_{m_{1}}^{l_{1}}(\lambda _{1})).e_{m_{2}}^{l_{2}}(\lambda _{2})=[%
\mathit{T}(e_{m_{1}}^{l_{1}}(\lambda
_{1}))]_{m_{3}m_{2}}.e_{m_{3}}^{l_{3}}(\lambda _{3})
\end{equation*}%
Comparing two last equations we get the statement of the theorem.
\end{proof}

Thus for the quantum superalgebra $U_{q}[osp(1\mid 2)]$ the Wigner-Eckart
theorem has exactly the same form as in the classical case $su(2)$ and
deformed case $U_{q}[su(2)].$ It is quite remarkable result because in
general all formulae in Racah-Wigner calculus for the quantum superalgebra $%
U_{q}[osp(1\mid 2)],$ although has similar form to corresponding formulae in
Racah-Wigner calculus for $su(2)$ and $U_{q}[su(2)],$ differ from the latter
by sometimes complicated phases \cite{2, 3}. We have avoided the appearance
of the not coventional phase in the Wigner-Eckart theorem using the modified
C-Gc.

The irreducible tensor operator $\mathit{T}$ for $U_{q}[osp(1\mid
2)]$ is even so we have $\mid \mathit{T}(e_{m}^{l}(\lambda ))\mid
=\mid e_{m}^{l}(\lambda )\mid =l-m+\lambda $ ${\rm mod}(2)$ and we
may introduce notation $\mathit{T}(e_{m}^{l}(\lambda ))\equiv
\mathit{T}_{m}^{l}(\lambda ). $ Let us write the defining
relations (2.9-2.11) for the components of irreducible tensor
operator $\mathit{T}_{m}^{l}(\lambda )$
\begin{equation*}
(-1)^{l-m}([l-m][l+m+1]\gamma )^{\frac{1}{2}}\mathit{T}_{m+1}^{l}(\lambda )=
\end{equation*}%
\begin{equation*}
=\pi ^{l_{3}}(v_{+})\circ \mathit{T}_{m}^{l}(\lambda )\circ \pi
^{l_{2}}(q^{-H})-(-1)^{l-m+\lambda }q^{\frac{1}{2}}\pi ^{l_{3}}(q^{-H})\circ
\mathit{T}_{m}^{l}(\lambda )\circ \pi ^{l_{2}}(v_{+})
\end{equation*}%
\begin{equation*}
([l+m][l-m+1]\gamma )^{\frac{1}{2}}\mathit{T}_{m-1}^{l}(\lambda )=
\end{equation*}%
\begin{equation*}
=\pi ^{l_{3}}(v_{-})\circ \mathit{T}_{m}^{l}(\lambda )\circ \pi
^{l_{2}}(q^{-H})-(-1)^{l-m+\lambda }q^{-\frac{1}{2}}\pi
^{l_{3}}(q^{-H})\circ \mathit{T}_{m}^{l}(\lambda )\circ \pi ^{l_{2}}(v_{-})
\end{equation*}%
\begin{equation*}
\frac{m}{2}\mathit{T}_{m}^{l}(\lambda )=\pi ^{l_{3}}(H)\circ \mathit{T}%
_{m}^{l}(\lambda )-\mathit{T}_{m}^{l}(\lambda )\circ \pi ^{l_{2}}(H)
\end{equation*}%
The above formulae are very similar to defining relations satisfied by the
components of irreducible tensor operator for the Hopf algebra $U_{q}[su(2)]$
\cite{6, 7, 12, 13}. The difference is only in the phase factor and the
definition of the symbol $[n].$ In the limit $q\rightarrow 1$ , for $l-m=0%
{\rm mod}(2)$ we get
\begin{equation*}
\frac{1}{2}(l-m)^{\frac{1}{2}}\mathit{T}_{m+1}^{l}(\lambda )=\pi
^{l_{3}}(v_{+})\circ \mathit{T}_{m}^{l}(\lambda )-(-1)^{\lambda }\mathit{T}%
_{m}^{l}(\lambda )\circ \pi ^{l_{2}}(v_{+})
\end{equation*}%
\begin{equation*}
\frac{1}{2}(l+m)^{\frac{1}{2}}\mathit{T}_{m-1}^{l}(\lambda )=\pi
^{l_{3}}(v_{-})\circ \mathit{T}_{m_{1}}^{l_{{}}}(\lambda )-(-1)^{\lambda }%
\mathit{T}_{m}^{l}(\lambda )\circ \pi ^{l_{2}}(v_{-})
\end{equation*}%
\begin{equation*}
\frac{m}{2}\mathit{T}_{m}^{l}(\lambda )=\pi ^{l_{3}}(H)\circ \mathit{T}%
_{m}^{l}(\lambda )-\mathit{T}_{m}^{l}(\lambda )\circ \pi ^{l_{2}}(H)
\end{equation*}%
and for $l-m=1\,{\rm mod}(2)$ we have
\begin{equation*}
-\frac{1}{2}(l+m+1)^{\frac{1}{2}}\mathit{T}_{m+1}^{l}(\lambda )=\pi
^{l_{3}}(v_{+})\circ \mathit{T}_{m}^{l}(\lambda )-(-1)^{\lambda }\mathit{T}%
_{m}^{l}(\lambda )\circ \pi ^{l_{2}}(v_{+})
\end{equation*}%
\begin{equation*}
\frac{1}{2}(l-m+1)^{\frac{1}{2}}\mathit{T}_{m-1}^{l}(\lambda )=\pi
^{l_{3}}(v_{-})\circ \mathit{T}_{m}^{l}(\lambda )-(-1)^{\lambda }\mathit{T}%
_{m}^{l}(\lambda )\circ \pi ^{l_{2}}(v_{-})
\end{equation*}%
\begin{equation*}
\frac{m}{2}\mathit{T}_{m}^{l}(\lambda )=\pi ^{l_{3}}(H)\circ \mathit{T}%
_{m}^{l}(\lambda _{1})-\mathit{T}_{m}^{l}(\lambda _{1})\circ \pi ^{l_{2}}(H)
\end{equation*}%
The above equations one can interpreted as defining relations for the
components of irreducible tensor operator for the Lie superlagebra $%
osp(1\mid 2).$ It is known that the Lie algebra $sl(2)$ generated by
elements $H,L_{\pm }=\pm 2[v_{\pm },v_{\pm }]_{+}$ is included in the
superalgebra $osp(1\mid 2)$ and we have
\begin{equation*}
\lbrack H,L_{\pm }]=\pm L_{\pm };[l_{+},L_{-}]=2H.
\end{equation*}%
Using the defining relations (2.8)\ for \textrm{a}$=H,L_{\pm }$ we get in
the limit $q\rightarrow 1$ the following equations
\begin{equation*}
-\frac{1}{4}\sqrt{(l-m)(l+m+2)}\mathit{T}_{m+2}^{l}(\lambda _{1})=\pi
^{l_{3}}(L_{+})\circ \mathit{T}_{m}^{l}(\lambda _{1})-\mathit{T}%
_{m}^{l}(\lambda _{1})\circ \pi ^{l_{2}}(L_{+})
\end{equation*}%
\begin{equation*}
-\frac{1}{4}\sqrt{(l+m)(l-m+2)}\mathit{T}_{m-2}^{l}(\lambda _{1})=\pi
^{l_{3}}(L_{-})\circ \mathit{T}_{m}^{l}(\lambda _{1})-\mathit{T}%
_{m}^{l}(\lambda _{1})\circ \pi ^{l_{2}}(L_{-})
\end{equation*}%
\begin{equation*}
\frac{m}{2}\mathit{T}_{m}^{l}(\lambda )=\pi ^{l_{3}}(H)\circ \mathit{T}%
_{m}^{l}(\lambda )-\mathit{T}_{m}^{l}(\lambda )\circ \pi ^{l_{2}}(H)
\end{equation*}%
for $l-m=0\,{\rm mod}(2)$ and
\begin{equation*}
-\frac{1}{4}\sqrt{(l-m-1)(l+m+1)}\mathit{T}_{m+2}^{l}(\lambda )=\pi
^{l_{3}}(L_{+})\circ \mathit{T}_{m}^{l}(\lambda _{1})-\mathit{T}%
_{m}^{l}(\lambda _{1})\circ \pi ^{l_{2}}(L_{+})
\end{equation*}%
\begin{equation*}
-\frac{1}{4}\sqrt{(l+m-1)(l-m+1)}\mathit{T}_{m-2}^{l}(\lambda )=\pi
^{l_{3}}(L_{-})\circ \mathit{T}_{m}^{l}(\lambda )-\mathit{T}_{m}^{l}(\lambda
)\circ \pi ^{l_{2}}(L_{-})
\end{equation*}%
\begin{equation*}
\frac{m}{2}\mathit{T}_{m}^{l}(\lambda )=\pi ^{l_{3}}(H)\circ \mathit{T}%
_{m}^{l}(\lambda )-\mathit{T}_{m}^{l}(\lambda )\circ \pi ^{l_{2}}(H)
\end{equation*}%
where $l-m=1\,{\rm mod}(2).$

These formulae are classical, Racah definition for components of irreducible
tensor operator for the Lie algebra $sl(2).$ Thus in the formal limit $%
U_{q}[osp(1\mid 2)]\rightarrow osp(1\mid 2)$ the set of the components $%
\mathit{T}_{m}^{l}(\lambda )$ of irreducible tensor operator
$\mathit{T}$ splits into two sets $\{\mathit{T}_{m}^{l}(\lambda
):l-m=0\,{\rm mod}(2)\}$ and $\{\mathit{T}_{m}^{l}(\lambda
):l-m=1\,{\rm mod}(2)\}$ which are sets of
components of irreducible tensor operators $\mathit{T}$ $^{l}$ and $\mathit{T%
}$ $^{l-1}$ for the Lie subalgebra $sl(2).$ Note that the sets $\{\mathit{T}%
_{m}^{l}(\lambda ):l-m=0\,{\rm mod}(2)\}$ and
$\{\mathit{T}_{m}^{l}(\lambda
):l-m=1\,{\rm mod}(2)\}$ differ in degree because we have $\mid \mathit{T}%
(e_{m}^{l}(\lambda ))\mid =l-m+\lambda $ $\,{\rm mod}(2).$ This
splitting is not surprising because the components of an
irreducible tensor operator has the same transformation rule as
the basis vectores of the irreducible representation. On the other
hand it is known that, with respect to $sl(2)$
a graded representation space $V^{l}$ of irreducible representation of $%
osp(1\mid 2)$ is a direct sum of two subspaces
\begin{equation*}
V^{l}=D^{l}(\lambda )\oplus D^{l-1}(\lambda +1)
\end{equation*}%
where $D^{l}(\lambda )$ and $D^{l-1}(\lambda +1)$ are the irreducible
representation spaces of the Lie algebra $sl(2).$ Thus our general
definition of tensor operators for $\mathtt{Z}_{2}$-graded Hopf in case of $%
U_{q}[osp(1\mid 2)],$ in the limit $q\rightarrow 1$ leads to classical
definition of tensor operators for the Lie algebra $sl(2)\subset osp(1\mid
2).$

From Wigner-Eckart theorem it follows that it is sufficient to know one
particular value of matrix element $[\mathit{T}_{m}^{l}(\lambda )]_{pq}$ of
\ tensor operator component $\mathit{T}_{m}^{l}(\lambda )$ to determine the
reduced matrix element $\alpha $ and then to express all remaining matrix
elements $[\mathit{T}_{m}^{l}(\lambda )]_{pq}$ in terms of Clebsch-Gordan
coefficients. It will be applied in the next section.

\section{Applications of Wigner-Eckart theorem.}

In this section we will consider tensor operators for the quantum
superalgebra $U_{q}[osp(1\mid 2)].$ First we construct in $U_{q}[osp(1\mid
2)]$ irreducible representations of highest weight $l$ (even natural number)
which will be irreducible subrepresentations of adjoint representation $%
(U_{q}[osp(1\mid 2)],ad)$.

\begin{proposition}
Let us define for any even natural $l$
\begin{equation*}
t_{m}^{l}=\left( \frac{[l+m]!}{[2l]![l-m]!}\right) ^{\frac{1}{2}%
}adv_{-}^{l-m}.v_{+}^{l}q^{lH}
\end{equation*}%
where $-l\leq m\leq l.$ Then
\begin{equation}
ade.t_{m}^{l}=([l-m][l+m+1])^{\frac{1}{2}}t_{m+1}^{l}
\end{equation}%
\begin{equation}
adf.t_{m}^{l}=([l+m][l-m+1])^{\frac{1}{2}}t_{m-1}^{l}
\end{equation}%
\begin{equation}
adH.t_{m}^{l}=\frac{m}{2}t_{m}^{l}.
\end{equation}%
We have also $\mid t_{l}^{l}\mid =\lambda =l=0$ $\,{\rm mod}(2)$
and $\mid t_{m}^{l}\mid =m$ $\,{\rm mod}(2).$ Therefore the
vectors $t_{m}^{l}$ form a basis of irreducible representation
$(U^{l},ad)$ of $U_{q}[osp(1\mid 2)]$ where $U^{l}\subset
U_{q}[osp(1\mid 2)].$
\end{proposition}

\begin{proof}
A direct calculation shows that $t_{l}^{l}$ is a highest weight vector of
weight $\frac{l}{2}$. The applying the standard procedure of construction of
the irreducible highest weight modul of $U_{q}[osp(1\mid 2)]$ we get the
result.
\end{proof}

\begin{corollary}
\bigskip The elements $t_{m}^{l}\in U_{q}[osp(1\mid 2)]$ are components of
the tensor operator $L^{l}\in Hom_{U_{q}[osp(1\mid 2)]}($ $%
U_{ad}^{l},Hom(U_{q}[osp(1\mid 2)]_{L},U_{q}[osp(1\mid 2)]_{L})).$
\end{corollary}

\begin{proof}
\bigskip The left regular action $L:U_{q}[osp(1\mid 2)]_{ad}\rightarrow
Hom(U_{q}[osp(1\mid 2)]_{L},U_{q}[osp(1\mid 2)]_{L})$ is a tensor operator
(Example 7, 9) and $U^{l}$ is an irreducible subrepresentation of $%
U_{q}[osp(1\mid 2)].$ So it is obvious that $L^{l}:U_{ad}^{l}\rightarrow
Hom(U_{q}[osp(1\mid 2)]_{L},U_{q}[osp(1\mid 2)]_{L})$ is also a tensor
operator. The equations (4.1-3) show that the components $t_{m}^{l}$ of $%
L^{l}$ satisfie the defining equation (2.8).
\end{proof}

As an application of the Wigner-Eckart theorem we will calculate the
matrices $\pi ^{j}(t_{m}^{l})_{pn}\equiv \lbrack t_{m}^{l}(j)]_{pn}$ of the
basis vectors $t_{m}^{l}$ of $(U^{l},ad)$ in the representation $%
(V^{j}(\lambda ),\pi ^{j}).$ Using the defining commutation relations for $%
U_{q}[osp(1\mid 2)]$ one can show that $t_{m}^{l}$ are rather complicated
combination of elements $H,$ $v_{\pm }$%
\begin{eqnarray}
t_{m}^{l} &=&\left( \frac{[l+m]!}{[2l]![l-m]!}\right) ^{\frac{1}{2}%
}\sum_{k}^{l-m}\sum_{p=0}^{N}(-1)^{\frac{k(k+1)}{2}}(-1)^{\frac{p(p-1)}{2}%
}q^{-\frac{k}{2}(l+m+1)}\frac{[l]![k]!}{[p]![l-p]![k-p]!}\times  \notag \\
&&\ \ \ \ \ \times %\QOVERD[ ]
\left[
 {l-m}\atop{k} \right]\gamma ^{p}v_{-}^{l-m}v_{+}^{l}\frac{[4H-k+l]!}{%
[4H-k+l-p]!}q^{mH}.
\end{eqnarray}%
where $N=\min (l,k)$ and we use a symbolic notation
\begin{equation*}
\frac{\lbrack H+m+p]!}{[H+m]!}\equiv \lbrack H+m+p]...[H+m+1].
\end{equation*}%
So a direct calculation of $\pi ^{j}(t_{m}^{l})_{pn}$ using matrices $\pi
^{j}(v_{\pm })_{mn}$, $\pi ^{j}(H)_{mn}$ seems to be difficult in general
case. However due to Wigner-Eckart theorem it is not necessary to do it. In
fact we have

\begin{theorem}
The basis vectors $t_{m}^{l}$ of $(U^{l},ad)$ have the following matrix form
in the irreducible representation $(V^{j}(\lambda ),\pi ^{j})$
\begin{equation*}
\pi ^{j}(t_{p}^{l})_{mn}=\alpha \lbrack lp0,jn\lambda \mid jm\lambda ]_{q}
\end{equation*}%
where
\begin{equation*}
\alpha =(-1)^{\frac{1}{2}l(l+1)}q^{-^{\frac{1}{2}l(l+1})}[l]!\left( \frac{%
[2j+l+1]!}{[2l]![2j-l]![2j+1]!}\gamma ^{l}\right) ^{\frac{1}{2}}
\end{equation*}%
is a reduced matrix element of the irreducible tensor operator $\pi
^{j}:U^{l}\rightarrow Hom(V^{j}(\lambda ),V^{j}(\lambda )).$

\begin{proof}
The representation $\pi ^{j}:U_{q}[osp(1\mid 2)]\rightarrow
Hom(V^{j}(\lambda ),V^{j}(\lambda ))$ is itself a tensor operator (Examples
6, 8). Because $U^{l}$ is an irreducible subrepersentation of $%
U_{q}[osp(1\mid 2)]$ then that $\pi ^{j}:U^{l}\rightarrow Hom(V^{j}(\lambda
),V^{j}(\lambda ))$ is an irreducible tensor opereator. Thus according to
the Wigner-Eckart theorem we have the following expression for matrix
element of components $\pi ^{j}(t_{p}^{l})$ of $\pi ^{j}$
\begin{equation*}
\pi ^{j}(t_{p}^{l})_{mn}=\alpha \lbrack lp0,jn\lambda \mid jm\lambda ]_{q}
\end{equation*}%
and in particular
\begin{equation}
\pi ^{j}(t_{l}^{l})_{mn}=\alpha \lbrack ll0,jn\lambda \mid jm\lambda ]_{q}.
\end{equation}%
Now on one hand from (3.1-3) we have
\begin{equation*}
\pi ^{j}(t_{l}^{l})_{mn}=(-1)^{\frac{1}{2}l(l+1)+l(j-m+l)}\left( \frac{%
[j-m+l]![j+m]!}{[j+m-l]![j-m]!}\gamma ^{l}\right) ^{\frac{1}{2}}q^{\frac{1}{2%
}l(m-l)}\delta _{mn+l}
\end{equation*}%
and on the other we have \cite{2}%
\begin{eqnarray*}
\lbrack ll0,jn\lambda &\mid &jm\lambda ]_{q}=q^{-\frac{n}{2}}q^{\frac{1}{4}%
(2j-l)(l+1)-\frac{1}{2}(j-m)(l+1)}\times \\
&&\times \left( \lbrack 2j+1]\frac{[2l]![2j-l]![j+m]![j-m+l]!}{%
[2j+l+1]![l]![l]![j-m]![j+m-l]!}\right) ^{\frac{1}{2}}\delta _{mn+l}.
\end{eqnarray*}%
After substitution of two last equations to equation (4.5 ) we get the value
of $\alpha .$
\end{proof}
\end{theorem}

At the end of this paper we give a method of constructing some elements of
the center of $U_{q}[osp(1\mid 2)]$ by use of the particular C-Gc $%
C_{mn}^{l}(\lambda )$ (3.6) and the elements $t_{p}^{l}$ of $U_{q}[osp(1\mid
2)].$ It is known that $C_{mn}^{l}(\lambda )$ couple two irreducible
representations $(V^{j},\pi ^{j})$ and $(V^{i},\pi ^{i})$ to one-dimensional
trivial representation. Therefore for any two irreducible representations $%
(U^{j},ad)$ and $(U^{i},ad)$ with bases $\{t_{m}^{j}\}$ and $\{t_{n}^{i}\},$
the following element $\mathfrak{C}^{j}$ of $U_{q}[osp(1\mid 2)]$
\begin{equation*}
\mathfrak{C}^{j}\mathfrak{=}\sum_{mn}(jm\lambda _{j},in\lambda _{i}\mid
00)_{q}t_{m}^{j}t_{n}^{i}
\end{equation*}%
form one dimensional trivial representation $(\mathfrak{C}^{j},\varepsilon )$
described in Example 5. It means that $\mathfrak{C}^{j}\in U_{q}[osp(1\mid
2)]_{\varepsilon }$ and consequently, from Proposition 1 it belongs to the
center of $U_{q}[osp(1\mid 2)]$.

\end{document}